\documentclass[journal,10pt]{IEEEtran}

\usepackage{hyperref}
\usepackage{cite}
\usepackage{graphicx}
\usepackage{array}
\usepackage{mdwmath}
\usepackage{amssymb}
\usepackage{mdwtab}
\usepackage{stfloats}
\usepackage[tight,footnotesize]{subfigure}
\usepackage{amsmath,amsthm,amsfonts}
\usepackage{threeparttable}
\usepackage{color}
\usepackage{url}
\usepackage{algorithm}
\usepackage{algorithmic}
\usepackage{multicol}
\usepackage{epstopdf}
\usepackage{flushend}
\usepackage{epsfig}
\usepackage{xcolor}
\usepackage{textcomp}
\usepackage{bm}
\usepackage{setspace}
\usepackage{float}

\ifCLASSINFOpdf

\else

\fi

\begin{document}
\title{Movable Antenna Enabled Interference Network: Joint Antenna Position and Beamforming Design}
\author{
	Honghao~Wang,
	Qingqing~Wu,~\IEEEmembership{Senior~Member,~IEEE,}
	and~Wen~Chen,~\IEEEmembership{Senior~Member,~IEEE}\vspace{-5pt}
	\thanks{H. Wang is with the Department of Electronic Engineering, Shanghai Jiao Tong University, Shanghai 200240, China, and also with the Department of Electrical and Computer Engineering, University of Macau, Macau 999078, China (e-mail: mc25018@um.edu.mo). Q. Wu and W. Chen are with the Department of Electronic Engineering, Shanghai Jiao Tong University, Shanghai 200240, China (e-mail: qingqingwu@sjtu.edu.cn; wenchen@sjtu.edu.cn).}
}

\markboth{}
{}

\maketitle

\begin{abstract}
This paper investigates the utility of movable antenna (MA) assistance for the multiple-input single-output (MISO) interference channel. We exploit an additional design degree of freedom provided by MA to enhance the desired signal and suppress interference so as to reduce the total transmit power of interference network. To this end, we jointly optimize the MA positions and transmit beamforming, subject to the signal-to-interference-plus-noise ratio constraints of users. To address the non-convex optimization problem, we propose an efficient iterative algorithm to alternately optimize the MA positions via successive convex approximation method and the transmit beamforming via second-order cone program approach. Numerical results demonstrate that the proposed MA-enabled MISO interference network outperforms its conventional counterpart without MA, which significantly enhances the capability of inter-cell frequency reuse and reduces the complexity of transmitter design.
\end{abstract}


\section{Introduction}\label{sec_intro}
Spectrum sharing is extensively employed in contemporary cellular networks with increasing node density, which renders the overall network performance limited by the co-channel interference. Consequently, more sophisticated techniques for interference management with multi-cell cooperation become essential. Existing works have attempted to leverage intelligent reflecting surface (IRS) or reconfigurable intelligent surface (RIS), which serves as a promising component for the next generation of wireless communication systems \cite{8811733}, to enlarge the achievable rate region of the multiple-input single-output (MISO) interference network \cite{9261117} and assist simultaneous wireless information and power transfer (SWIPT) in it \cite{10268916}. However, the antenna scheme implemented at transceivers is predominantly the conventional FPA, which restricts the performance of interference networks due to underutilization of channel variation in the continuous spatial field.

To harness additional spatial degrees of freedom (DoFs) in wireless channels for enhancing the diversity and spatial multiplexing of wireless systems, movable antenna (MA) \cite{10318061} or fluid antenna system (FAS) \cite{10146274} was conceived, which is designed to overcome the constraints of conventional fixed-position antennas (FPAs). Specifically, by connecting the MAs to radio frequency (RF) chains via flexible cables, the MA positions are allowed to be adjusted over a two-dimensional (2D) region through motors or servos \cite{8060521}. Based on this, MAs can be timely deployed to the positions with more favorable channel conditions for improving the quality-of-service.

With the aforementioned advantages, MA has garnered significant attention \cite{10243545,10354003,10416896,10416363,10414081,10430366,10388242}. For instance, to further improve the channel capacity of multiple-input multiple-output (MIMO) system, authors in \cite{10243545} proposed a new architecture incorporating MA, which demonstrates a substantial enhancement in communication performance compared to the conventional FPA system. Moreover, an MA-enabled multi-access channel for multi-user uplink transmission was investigated in \cite{10354003}, which leads to a noteworthy reduction in total transmit power as opposed to the FPA systems.

In light of the above, this paper investigates the MISO interference channel aided by MAs at transmitters, leveraging the inherent characteristics of the multi-path channel and the extra spatial DoFs it offers to reduce inter-cell interference. The integration of MA introduces a new DoF in system design, enabling both desired signal enhancement and interference mitigation. To this end, by jointly optimizing MA positions and transmit beamforming, we aim to minimize the total transmit power under the individual signal-to-interference-plus-noise ratio (SINR) requirement of each user, which is a highly coupled non-convex problem. To address this challenge, an efficient algorithm based on block coordinate descent (BCD) is proposed, where MA positions and beamforming vectors are optimized in an alternating manner. Specifically, with fixed MA positions, the optimal beamforming vectors are obtained by second-order cone program (SOCP). On the other hand, by fixing beamforming vectors and scaling SINR constraints with a meticulously designed auxiliary value, the MA positions can be updated iteratively via successive convex approximation (SCA). Simulation results validate that the proposed algorithm can be effectively utilized in the MISO interference network, provided a certain region size for antennas moving is available. Consequently, the performance of spectrum sharing in interference network is dramatically improved. The MA system with simple beamforming, e.g., maximum ratio transmission (MRT), performs only slightly worse than that with complex beamforming and significantly better than the FPA system. Moreover, the number of antennas required for MA-aided interference network is drastically reduced, thus enabling the simplification of transmitter design.

\emph{Notations:} $\mathbf{x}\!\left(n\right)$, $\mathbf{X}\!\left(m,n\right)$, $\left|\left|\mathbf{X}\right|\right|_F$, $\lambda_{\text{max}}\!\left\{\mathbf{X}\right\}$, and $\operatorname{vec}\!\left(\mathbf{X}\right)$ represent the $n$th entry of vector $\mathbf{x}$, the $\left(m,n\right)$th element of matrix $\mathbf{X}$, the Frobenius norm of $\mathbf{X}$, the maximum eigenvalue of $\mathbf{X}$, and the vectorization of $\mathbf{X}$, respectively. $\operatorname{diag}\!\left(\mathbf{x}\right)$ denotes the diagonal matrix with diagonal elements being $\mathbf{x}$.

\section{System Model and Problem Formulation}\label{sec_model}
\begin{figure}[t]
	\centering
	\includegraphics[width=2.45in]{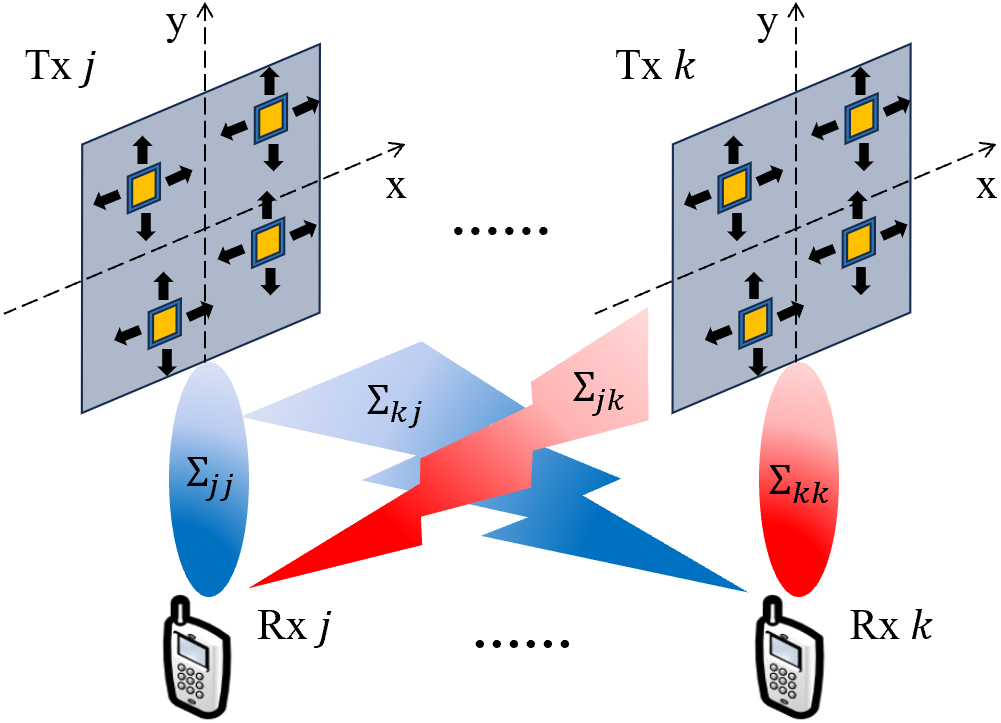}
	\caption{A MISO interference network aided by MAs.}
	\label{fig1_model}
	\vspace{-5pt}
\end{figure}

As shown in Fig. \ref{fig1_model}, we consider a MISO interference channel with $K$ transmitter-user pairs, where each transmitter equipped with $N$ MAs simultaneously transmits signals to its corresponding single-antenna user. Specifically, the MAs are connected to RF chains via flexible cables, enabling real-time adjustment of their positions \cite{8060521}. The collections of coordinates of $N$ MAs in each transmitter are denoted as $\mathbf{T}_j=\left[\mathbf{t}_{j,1},\ldots,\mathbf{t}_{j,N}\right]\in\mathbb{R}^{2{\times}N},~j\in\left\{1,\ldots,K\right\}$, where $\mathbf{t}_{j,n}=\left[x_{j,n},y_{j,n}\right]^T\in\mathcal{C}_j$ for $n\in\left\{1,\ldots,N\right\}$ represented by Cartesian coordinates indicates the position of the $n$th MA in transmitter $j$. Without loss of generality, $\mathcal{C}_j$ can be set as 2D square regions with the identical size of $A\times{A}$ wherein the MAs can move without restraint.

According to the field-response model \cite{10318061}, the channel vector $\mathbf{h}_{kj}\in\mathbb{C}^{N\times1}$ from transmitter $j$ to user $k\in\left\{1,\ldots,K\right\}$ follows the structure as
\begin{equation}\label{eq_channel_1}
	\mathbf{h}_{kj}=\mathbf{G}_{kj}^H\mathbf{\Sigma}_{kj}\mathbf{1},
\end{equation}
where $\mathbf{1}\in\left\{1\right\}^{L_{kj}\times1}$ is the all-one field response vector (FRV) and $L_{kj}$ is the number of channel paths from transmitter $j$ to user $k$. The path-response matrix (PRM) is defined as $\mathbf{\Sigma}_{kj}=\operatorname{diag}\left(\left[\tau_{kj,1},\ldots,\tau_{kj,L_{kj}}\right]^T\right)\in\mathbb{C}^{L_{kj}\times{L_{kj}}}$, where $\tau_{kj,l}$ is the complex response of the $l$th path for $l=1,\ldots,L_{kj}$. $\mathbf{G}_{kj}=\left[\mathbf{g}_{kj,1},\ldots,\mathbf{g}_{kj,N}\right]\in\mathbb{C}^{L_{kj}\times{N}}$ is the transmit field response matrix (FRM) at transmitter $j$, where $\mathbf{g}_{kj,n}\in\mathbb{C}^{L_{kj}\times1}$ is the transmit FRV between user $k$ and the $n$th MA in transmitter $j$ for $n=1,\ldots,N$. Denote $\lambda$ as the carrier wavelength, the transmit FRVs are further given by
\begin{equation}\label{eq_FRV}
	\mathbf{g}_{kj,n}=\left[e^{j\frac{2\pi}{\lambda}\mathbf{t}_{j,n}^T\mathbf{p}_{kj,1}},\ldots,e^{j\frac{2\pi}{\lambda}\mathbf{t}_{j,n}^T\mathbf{p}_{kj,L_{kj}}}\right]^T,
\end{equation}
where $\mathbf{p}_{kj,l}=\left[\sin\theta_{kj,l}\cos\phi_{kj,l},\cos\theta_{kj,l}\right]^T$, $\theta_{kj,l}\in\left[0,\pi\right]$ and $\phi_{kj,l}\in\left[0,\pi\right]$ denote the elevation and azimuth angles of the $l$th path from transmitter $j$ to user $k$, respectively. Moreover, the channel vectors can be reformulated as $\mathbf{h}_{kj}=\left[h_{kj}\left(\mathbf{t}_{j,1}\right),\ldots,h_{kj}\left(\mathbf{t}_{j,N}\right)\right]^T$, where
\begin{equation}\label{eq_chan_re}
	h_{kj}\left(\mathbf{t}_{j,n}\right)\overset{\triangle}{=}\sum_{l=1}^{L_{kj}}\tau_{kj,l}e^{-j\frac{2\pi}{\lambda}\mathbf{t}_{j,n}^T\mathbf{p}_{kj,l}}.
\end{equation}

Let $\mathbf{w}_j\in\mathbb{C}^{N\times1}$ denote the transmit beamforming vector of transmitter $j$. Then, the baseband complex signal received at user $k$ can be expressed as
\begin{equation}\label{eq_rec}
	\!\!\!y_k\!=\!\sum_{j=1}^{K}\mathbf{h}_{kj}^H\mathbf{w}_{j}s_j\!+\!z_k\!=\!\underbrace{\vphantom{\sum_{j\ne{k}}\mathbf{h}_{kj}^H\mathbf{w}_{j}s_j}\mathbf{h}_{kk}^H\mathbf{w}_ks_k}_{\text{desired signal}}\!+\!\!\underbrace{\vphantom{\sum_{j\ne{k}}\mathbf{h}_{kj}^H\mathbf{w}_{j}s_j}\sum_{j\ne{k}}\mathbf{h}_{kj}^H\mathbf{w}_{j}s_j}_{\text{interference}}\!+z_k,~\forall{k},
\end{equation}
where $s_k$ is the information symbol for user $k$ with normalized power, $z_k$ denotes the additive white Gaussian noise (AWGN) at user $k$, which is assumed to be circularly symmetric complex Gaussian (CSCG) distributed with zero mean and power $\sigma_k^2$, i.e., $z_k\sim\mathcal{CN}\left(0,\sigma_k^2\right)$. With interference treated as noise at each user and signaling assumed CSCG, the resulting SINR at user $k$ is given by
\begin{equation}\label{eq_rate}
	\gamma_k=\frac{\left|\mathbf{h}_{kk}^H\mathbf{w}_k\right|^2}{\sum_{j\ne{k}}\left|\mathbf{h}_{kj}^H\mathbf{w}_j\right|^2+\sigma_k^2},~\forall{k}.
\end{equation}

In this paper, we aim to minimize the total transmit power by jointly optimizing the MA positions $\left\{\mathbf{T}_j\right\}_{j=1}^K$ and beamforming vectors $\left\{\mathbf{w}_j\right\}_{j=1}^K$, subject to the individual SINR constraint of each user. Besides, to avoid coupling effect between MAs in transmit regions, a minimum distance $D$ is required between each pair of MAs in the same transmitter, i.e., $\left|\left|\mathbf{t}_{j,n}-\mathbf{t}_{j,\tilde{n}}\right|\right|{\ge}D,~{\forall}j,n,\tilde{n},\tilde{n}{\ne}n,~\tilde{n}\in\left\{1,\ldots,N\right\}$. Accordingly, the optimization problem is formulated as
\begin{subequations}\label{eq_P1}
	\begin{alignat}{2}
		\!\!\text{(P1)}:&\underset{\left\{\mathbf{w}_j\right\}_{j=1}^K,\left\{\mathbf{T}_j\right\}_{j=1}^K}{\min}~&&\sum_{j=1}^{K}\left|\left|\mathbf{w}_j\right|\right|^2\label{eq_P1_a}\\
		&~\quad\quad\mathrm{s.t.}&&\gamma_k\ge\gamma_{k,\text{min}},~\forall{k},\label{eq_P1_b}\\
		&&&\mathbf{t}_{j,n}\in\mathcal{C}_j,~{\forall}j,n,\label{eq_P1_c}\\
		&&&\!\left|\left|\mathbf{t}_{j,n}-\mathbf{t}_{j,\tilde{n}}\right|\right|{\ge}D,~{\forall}j,n,\tilde{n},\tilde{n}{\ne}n,\label{eq_P1_d}
	\end{alignat}
\end{subequations}
where $\gamma_{k,\text{min}}$ is the minimum SINR requirement for user $k$. Note that (P1) is a non-convex optimization problem due to the non-convexity of minimum SINR constraint in (\ref{eq_P1_b}) as well as minimum distance constraint in (\ref{eq_P1_d}), which makes (P1) challenging to address.

\section{Proposed Solution}\label{sec_alg}
In this section, by dividing problem (P1) into two subproblems (P2) and (P3), an alternating optimization (AO) algorithm based on the SOCP and SCA techniques is proposed, where the transmit beamforming and MA positions are updated alternately with the other fixed. First, we consider the subproblem of transmit beamforming optimization with given MA positions. In this case, problem (P1) is transformed into
\begin{subequations}\label{eq_P2}
	\begin{alignat}{2}
			\text{(P2)}:~&\underset{\left\{\mathbf{w}_j\right\}_{j=1}^K}{\min}~&&\sum_{j=1}^{K}\left|\left|\mathbf{w}_j\right|\right|^2\label{eq_P2_a}\\
			&~~\mathrm{s.t.}&&~\!\text{(\ref{eq_P1_b})}.\nonumber
		\end{alignat}
\end{subequations}
Problem (P2) is reduced to the transmit beamforming optimization problem of the conventional MISO interference channel, which can be optimally solved via SOCP together with the bisection search method \cite{6170850,8910627}.

\begin{figure*}[!t]
	\normalsize
	\small
	\newcounter{al1}
	\setcounter{al1}{\value{equation}}
	\setcounter{equation}{11}
	\begin{align}
		&f_{kj}\left(\mathbf{t}_{j,n}^i\right)+{\nabla}f_{kj}^T\left(\mathbf{t}_{j,n}^i\right)\left(\mathbf{t}_{j,n}-\mathbf{t}_{j,n}^i\right)-\frac{\delta_{kj,n}}{2}\left(\mathbf{t}_{j,n}-\mathbf{t}_{j,n}^i\right)^T\left(\mathbf{t}_{j,n}-\mathbf{t}_{j,n}^i\right)~{\le}~f_{kj}\left(\mathbf{t}_{j,n}\right)\nonumber\\
		&\quad\quad\quad\quad\quad\quad\quad\quad\quad\quad\quad\quad\quad\quad\quad\quad\quad{\le}~f_{kj}\left(\mathbf{t}_{j,n}^i\right)+{\nabla}f_{kj}^T\left(\mathbf{t}_{j,n}^i\right)\left(\mathbf{t}_{j,n}-\mathbf{t}_{j,n}^i\right)+\frac{\delta_{kj,n}}{2}\left(\mathbf{t}_{j,n}-\mathbf{t}_{j,n}^i\right)^T\left(\mathbf{t}_{j,n}-\mathbf{t}_{j,n}^i\right).\label{eq_Taylor}\\[-4pt]
		&\frac{\partial^2f_{kj}}{\partial{x_{j,n}^2}}=-\frac{4\pi^2}{\lambda^2}\sum_{a=1}^{L_{kj}-1}\sum_{b=a+1}^{L_{kj}}\zeta_n\!\left(a,b,kj\right)\left(-\sin\theta_{kj,a}\cos\phi_{kj,a}+\sin\theta_{kj,b}\cos\phi_{kj,b}\right)^2-\frac{4\pi^2}{\lambda^2}\sum_{l=1}^{L_{kj}}\chi_n\!\left(kj,l\right)\sin^2\theta_{kj,l}\cos^2\phi_{kj,l}.\label{eq_partial_x}\\[-3pt]
		&\frac{\partial^2f_{kj}}{\partial{y_{j,n}^2}}=-\frac{4\pi^2}{\lambda^2}\sum_{a=1}^{L_{kj}-1}\sum_{b=a+1}^{L_{kj}}\zeta_n\!\left(a,b,kj\right)\left(-\cos\theta_{kj,a}+\cos\theta_{kj,b}\right)^2-\frac{4\pi^2}{\lambda^2}\sum_{l=1}^{L_{kj}}\chi_n\!\left(kj,l\right)\cos^2\theta_{kj,l}.\label{eq_partial_y}\\[-3pt]
		&\frac{\partial^2f_{kj}}{\partial{x_{j,n}}\partial{y_{j,n}}}=\frac{\partial^2f_{kj}}{\partial{y_{j,n}}\partial{x_{j,n}}}=-\frac{4\pi^2}{\lambda^2}\sum_{a=1}^{L_{kj}-1}\sum_{b=a+1}^{L_{kj}}\zeta_n\!\left(a,b,kj\right)\left(-\sin\theta_{kj,a}\cos\phi_{kj,a}+\sin\theta_{kj,b}\cos\phi_{kj,b}\right)\left(-\cos\theta_{kj,a}+\cos\theta_{kj,b}\right)\nonumber\\[-4pt]
		&~\quad\quad\quad\quad\quad\quad\quad\quad\quad\quad\quad\quad-\frac{4\pi^2}{\lambda^2}\sum_{l=1}^{L_{kj}}\chi_n\left(kj,l\right)\sin\theta_{kj,l}\cos\phi_{kj,l}\cos\theta_{kj,l}.\label{eq_partial_xy}\\[-6pt]
		&\lambda_{\text{max}}\left\{\nabla^2f_{kj}\!\left(\mathbf{t}_{j,n}\right)\right\}=\left|\left|\nabla^2f_{kj}\!\left(\mathbf{t}_{j,n}\right)\right|\right|^2\le\left|\left|\nabla^2f_{kj}\!\left(\mathbf{t}_{j,n}\right)\right|\right|_F^2=\left(\frac{\partial^2f_{kj}}{\partial{x_{j,n}^2}}\right)^2+\left(\frac{\partial^2f_{kj}}{\partial{x_{j,n}}\partial{y_{j,n}}}\right)^2+\left(\frac{\partial^2f_{kj}}{\partial{y_{j,n}}\partial{x_{j,n}}}\right)^2+\left(\frac{\partial^2f_{kj}}{\partial{y_{j,n}^2}}\right)^2\nonumber\\[-5pt]
		&\!~~\quad\quad\quad\quad\quad\quad\quad\quad\quad\quad\quad\quad\quad\quad\quad\quad\le4\left(\frac{4\pi^2}{\lambda^2}\sum_{a=1}^{L_{kj}-1}\sum_{b=a+1}^{L_{kj}}2\left|\mathbf{V}_{kj}\left(a,b\right)\right|+\frac{4\pi^2}{\lambda^2}\sum_{l=1}^{L_{kj}}2\left|\mathcal{G}_{kj,\tilde{n}}\right|\left|w_{j,n}\right|\left|v_{kj,l}\right|\right)^2.\label{eq_norm_inequ}\\[-1pt]
		\setcounter{equation}{17}
		&f_{kk}\left(\mathbf{t}_{k,n}^i\right)+{\nabla}f_{kk}^T\left(\mathbf{t}_{k,n}^i\right)\left(\mathbf{t}_{k,n}-\mathbf{t}_{k,n}^i\right)-\frac{\delta_{kk,n}}{2}\left(\mathbf{t}_{k,n}-\mathbf{t}_{k,n}^i\right)^T\left(\mathbf{t}_{k,n}-\mathbf{t}_{k,n}^i\right)\nonumber\\[-2pt]
		&\quad\quad\quad\quad\quad\!{\ge}~\gamma_{k,\text{min}}\left(\sum_{j\ne{k}}\left(f_{kj}\left(\mathbf{t}_{j,n}^i\right)+{\nabla}f_{kj}^T\left(\mathbf{t}_{j,n}^i\right)\left(\mathbf{t}_{j,n}-\mathbf{t}_{j,n}^i\right)+\frac{\delta_{kj,n}}{2}\left(\mathbf{t}_{j,n}-\mathbf{t}_{j,n}^i\right)^T\left(\mathbf{t}_{j,n}-\mathbf{t}_{j,n}^i\right)\right)+\sigma_k^2\right),~\forall{k}.\label{eq_SINR_SCA}
	\end{align}
	\hrulefill
	\setcounter{equation}{\value{al1}}
	\vspace{-5pt}
\end{figure*}

Next, for any given optimal $\left\{\mathbf{w}_j\right\}_{j=1}^K$, problem (P1) is transformed into the following feasible check problem (P3), where the MA positions $\left\{\mathbf{T}_j\right\}_{j=1}^K$ can be optimized with the constraints (\ref{eq_P1_b})-(\ref{eq_P1_d}).
\vspace{-3pt}
\begin{subequations}\label{eq_P3}
	\begin{alignat}{2}
		\text{(P3)}:~&\operatorname{Find}~&&\left\{\mathbf{T}_j\right\}_{j=1}^K\label{eq_P3_a}\\
		&~\mathrm{s.t.}&&~\text{(\ref{eq_P1_b}), (\ref{eq_P1_c}) and (\ref{eq_P1_d})}.\nonumber
	\end{alignat}
\end{subequations}
Due to that for any user $k$, the channel power gain between a transmitter and that user is solely affected by the MA positions of that transmitter, i.e., the variables $\left\{\mathbf{T}_j\right\}_{j=1}^K$ are mutually independent, we decompose (P3) into $N$ subproblems (P3.1.n), where the $n$th MA of each transmitter is optimized in parallel with all other MAs fixed in the $n$th subproblem.
\vspace{-2pt}
\begin{subequations}\label{eq_P3.1.n}
	\begin{alignat}{2}
		\text{(P3.1.n)}:~&\operatorname{Find}~&&\left\{\mathbf{t}_{j,n}\right\}_{j=1}^K\label{eq_P3.1.n_a}\\
		&~\mathrm{s.t.}&&\gamma_k\ge\gamma_{k,\text{min}},~\forall{k},\label{eq_P3.1.n_b}\\
		&&&\mathbf{t}_{j,n}\in\mathcal{C}_j,~{\forall}j,\label{eq_P3.1.n_c}\\
		&&&\!\left|\left|\mathbf{t}_{j,n}-\mathbf{t}_{j,\tilde{n}}\right|\right|{\ge}D,~\forall{j},\tilde{n},\tilde{n}\ne{n}.\label{eq_P3.1.n_d}
	\end{alignat}
\end{subequations}
By expanding the numerator and denominator of $\gamma_k$ in (\ref{eq_P3.1.n_b}), in terms of $\left|\mathbf{h}_{kj}^H\mathbf{w}_j\right|^2$, we have the following reconstructed equation:
\vspace{-2pt}
\begin{align}\label{eq_gain_re_1}
	&\!\!\!\!\left|\mathbf{h}_{kj}^H\mathbf{w}_j\right|^2\!=\!\left(w_{j,n}\mathbf{v}_{kj}^H\mathbf{g}_{kj,n}\!+\mathcal{G}_{kj,\tilde{n}}\right)\!\left(w_{j,n}^*\mathbf{g}_{kj,n}^H\mathbf{v}_{kj}\!+\mathcal{G}_{kj,\tilde{n}}^*\right)\nonumber\\
	&\!\!\!\!=\!\left|w_{j,n}\mathbf{v}_{kj}^H\mathbf{g}_{kj,n}\right|^2\!\!\!+\!2\!\operatorname{Re}\!\left\{\mathcal{G}_{kj,\tilde{n}}^*w_{j,n}\mathbf{v}_{kj}^H\mathbf{g}_{kj,n}\right\}\!+\!\left|\mathcal{G}_{kj,\tilde{n}}\right|^2,
\end{align}
where $\mathcal{G}_{kj,\tilde{n}}=\sum_{\tilde{n}{\ne}n}w_{j,\tilde{n}}\sum_{l=1}^{L_{kj}}\tau_{kj,l}^*e^{j\frac{2\pi}{\lambda}\mathbf{t}_{j,\tilde{n}}^T\mathbf{p}_{kj,l}}$, $\mathbf{v}_{kj}=\operatorname{vec}\left(\mathbf{\Sigma}_{kj}\right)$, $w_{j,n}=\mathbf{w}_j\left(n\right)$. Then, by denoting $f_{kj}\left(\mathbf{t}_{j,n}\right)=\left|\mathbf{h}_{kj}^H\mathbf{w}_j\right|^2$, (\ref{eq_gain_re_1}) can be further rewritten as
\vspace{-5pt}
\begin{align}\label{eq_gain_re_2}
	f_{kj}\left(\mathbf{t}_{j,n}\right)&=\operatorname{tr}\left(\mathbf{V}_{kj,n}\right)+\left|\mathcal{G}_{kj,\tilde{n}}\right|^2\nonumber\\[-2pt]
	&+\sum_{a=1}^{L_{kj}-1}\!\sum_{b=a+1}^{L_{kj}}\zeta_n\!\left(a,b,kj\right)+\sum_{l=1}^{L_{kj}}\chi_n\!\left(l,kj\right),
\end{align}
where we define
\vspace{-1pt}
\begin{align}
	&\mathbf{V}_{kj,n}\!\overset{\triangle}{=}\!\left|w_{j,n}\right|^2\!\mathbf{v}_{kj}\mathbf{v}_{kj}^H,~r_{kj,n,l}\!\overset{\triangle}{=}\!\left|\mathcal{G}_{kj,\tilde{n}}\right|\!\left|w_{j,n}\right|\!\left|v_{kj,l}\right|\!,\nonumber\\
	&\alpha_n\!\left(a,b,kj\right)\!\overset{\triangle}{=}\!\left(2\pi/\lambda\right)\!\left(-\mathbf{t}_{j,n}^T\mathbf{p}_{kj,a}\!+\!\mathbf{t}_{j,n}^T\mathbf{p}_{kj,b}\right)\!,~v_{kj,l}\!\overset{\triangle}{=}\!\mathbf{v}_{kj}\!\left(l\right)\!,\nonumber\\
	&\zeta_n\!\left(a,b,kj\right)\!\overset{\triangle}{=}\!2\!\left|\mathbf{V}_{kj,n}\!\left(a,b\right)\right|\cos\!\left(\alpha_n\!\left(a,b,kj\right)\!+\!\angle\mathbf{V}_{kj,n}\!\left(a,b\right)\right)\!,\nonumber\\
	&\chi_n\!\left(kj,l\right)\!\overset{\triangle}{=}\!2r_{kj,n,l}\cos\!\left(\!\frac{2\pi}{\lambda}\mathbf{t}_{j,n}^T\mathbf{p}_{kj,l}\!-\!\angle\mathcal{G}_{kj,\tilde{n}}\!+\!\angle{w_{j,n}}\!-\!\angle{v_{kj,l}}\!\right)\!.\nonumber
\end{align}

\vspace{-2pt}
Next, the SCA method is employed to tackle the non-convex constraint (\ref{eq_P3.1.n_b}) based on the formulations above. By applying the second-order Taylor expansion, we construct a quadratic surrogate function that serves as a global lower bound for $\gamma_k$ in (\ref{eq_P3.1.n_b}). Specifically, given the provided local point $\mathbf{t}_{j,n}^i$ obtained in the $i$th iteration, the inequality relationship (\ref{eq_Taylor}) holds as shown at the top of this page. Accordingly, the upper and lower bounds of $f_{kj}\!\left(\mathbf{t}_{j,n}\right)$ are obtained, which can be achieved through the introduction of a positive real number $\delta_{kj,n}$ such that $\delta_{kj,n}\mathbf{I}_2\succeq\nabla^2f_{kj}\!\left(\mathbf{t}_{j,n}\right)$. According to (\ref{eq_partial_x})-(\ref{eq_norm_inequ}), $\delta_{kj,n}$ can be selected by calculating the Frobenius norm of the Hessian matrix of $\nabla^2f_{kj}\!\left(\mathbf{t}_{j,n}\right)$ as follows:
\setcounter{equation}{16}
\vspace{-2pt}
\begin{equation}\label{eq_delta}
	\delta_{kj,n}=\frac{16\pi^2}{\lambda^2}\!\left(\sum_{a=1}^{L_{kj}-1}\!\sum_{b=a+1}^{L_{kj}}\!\left|\mathbf{V}_{kj}\left(a,b\right)\right|+\sum_{l=1}^{L_{kj}}r_{kj,n,l}\right).
	\vspace{-2pt}
\end{equation}
In this way, by tightening the numerator of $\gamma_k$ in (\ref{eq_P3.1.n_b}) to its lower bound and the denominator to its upper bound, (P3.1.n) is then reduced to problem (P3.2.n) in the $\left(i+1\right)$th iteration, where constraint (\ref{eq_SINR_SCA}) is shown at the top of this page.
\setcounter{equation}{18}
\begin{subequations}\label{eq_P3.2.n}
	\begin{alignat}{2}
		\text{(P3.2.n)}:~&\operatorname{Find}~&&\left\{\mathbf{t}_{j,n}\right\}_{j=1}^K\\
		&~\mathrm{s.t.}&&~\text{(\ref{eq_SINR_SCA}), (\ref{eq_P3.1.n_c}) and (\ref{eq_P3.1.n_d})}.\nonumber
	\end{alignat}
\end{subequations}

\vspace{-4pt}
Finally, we deal with the non-convex constraint (\ref{eq_P3.1.n_d}) via the SCA method. Denote gradient vector of $\left|\left|\mathbf{t}_{j,n}-\mathbf{t}_{j,\tilde{n}}\right|\right|_2$ over $\mathbf{t}_{j,n}$ as $\nabla\left|\left|\mathbf{t}_{j,n}-\mathbf{t}_{j,\tilde{n}}\right|\right|_2=\left(\mathbf{t}_{j,n}-\mathbf{t}_{j,\tilde{n}}\right)/\left|\left|\mathbf{t}_{j,n}-\mathbf{t}_{j,\tilde{n}}\right|\right|_2$. Since the denominator term $\left|\left|\mathbf{t}_{j,n}-\mathbf{t}_{j,\tilde{n}}\right|\right|_2\!\ge\!{D}\!>\!0$, the gradient vector always exists. Furthermore, the following inequality holds utilizing the first-order Taylor expansion at $\mathbf{t}_{j,n}^i$:
\begin{align}\label{eq_dis}
	&\!\left|\left|\mathbf{t}_{j,n}\!-\!\mathbf{t}_{j,\tilde{n}}\right|\right|\ge\left|\left|\mathbf{t}_{j,n}^i\!-\!\mathbf{t}_{j,\tilde{n}}\right|\right|\!+\!\left(\nabla\!\left|\left|\mathbf{t}_{j,n}^i\!-\!\mathbf{t}_{j,\tilde{n}}\right|\right|\right)^T\!\left(\mathbf{t}_{j,n}\!-\!\mathbf{t}_{j,\tilde{n}}\right)\nonumber\\[3pt]
	&\!=\left|\left|\mathbf{t}_{j,n}^i-\mathbf{t}_{j,\tilde{n}}\right|\right|+\frac{1}{\left|\left|\mathbf{t}_{j,n}^i-\mathbf{t}_{j,\tilde{n}}\right|\right|}\left(\mathbf{t}_{j,n}^i-\mathbf{t}_{j,\tilde{n}}\right)^T\!\left(\mathbf{t}_{j,n}-\mathbf{t}_{j,n}^i\right)\nonumber\\
	&\!=\frac{1}{\left|\left|\mathbf{t}_{j,n}^i-\mathbf{t}_{j,\tilde{n}}\right|\right|}\left(\mathbf{t}_{j,n}^i-\mathbf{t}_{j,\tilde{n}}\right)^T\left(\mathbf{t}_{j,n}-\mathbf{t}_{j,\tilde{n}}\right).
\end{align}
Hereto, for a given $\mathbf{t}_{j,n}^i$ obtained in the $i$th iteration, (P3.2.n) is evidently transformed into a convex position optimization problem (P3.3.n) in the $\left(i+1\right)$th iteration.
\begin{subequations}\label{eq_P3.3.n}
	\begin{alignat}{2}
		\text{(P3.3.n)}:~&\operatorname{Find}~&&\left\{\mathbf{t}_{j,n}\right\}_{j=1}^K\label{eq_P14_a}\\
		&\!\!\!\!\!\!\!\!\!\!\!\!\!\!\!\!\mathrm{s.t.}&&\!\!\!\!\!\!\!\!\!\!\!\!\!\!\!\!\!\!\!\!\frac{\left(\mathbf{t}_{j,n}^i\!-\!\mathbf{t}_{j,\tilde{n}}\right)^T\!\left(\mathbf{t}_{j,n}\!-\!\mathbf{t}_{j,\tilde{n}}\right)}{\left|\left|\mathbf{t}_{j,n}^i\!-\!\mathbf{t}_{j,\tilde{n}}\right|\right|}\!\ge\!{D},~\forall{j},\tilde{n},\tilde{n}\!\ne\!{n},\label{eq_P3.3.n_b}\\
		&&&\!\!\!\!\!\!\!\!\!\!\!\!\!\!\!\!\!\!\!\text{(\ref{eq_SINR_SCA}), (\ref{eq_P3.1.n_c})}.\nonumber
	\end{alignat}
\end{subequations}

Based on the above derivations, the proposed algorithm for joint antenna position and beamforming design is summarized in \textbf{Algorithm 1}, where the objective function is monotonically non-increasing and lower bounded, thereby ensuring the convergence. The complexities for solving (P2) and (P3.3.n) are $\mathcal{O}\left(K^{3.5}N^3\right)$ and $\mathcal{O}\left(K^{3.5}N^{1.5}\right)$, respectively. Therefore, the overall complexity of the proposed algorithm is $\mathcal{O}\left(I_1\left(K^{3.5}N^3+NI_2\left(K^{3.5}N^{1.5}\right)\right)\right)$, where $I_1$ and $I_2$ denote the numbers of iterations.

\begin{algorithm}[htbp]
	\renewcommand{\algorithmicrequire}{\textbf{Input:}}
	\renewcommand{\algorithmicensure}{\textbf{Output:}}
	\caption{Joint beamforming and antenna position design}
	\begin{algorithmic}[1]
		\label{alg_1}
		\REQUIRE Convergence criteria and initial positions $\left\{\mathbf{T}_j^{\left[0\right]}\right\}_{j=1}^K$.
		\ENSURE The optimized solution $\left\{\left\{\mathbf{w}_j^\star\right\}_{j=1}^K,\left\{\mathbf{T}_j^\star\right\}_{j=1}^K\right\}$.
		\REPEAT
		\STATE Update $\left\{\mathbf{w}_j^\star\right\}_{j=1}^K$ by solving (P2).
		\STATE Update $\left\{\mathbf{T}_j^\star\right\}_{j=1}^K$ by solving (P3.3.n).
		\UNTIL{The objective function of (P1) converges.}
	\end{algorithmic}
\end{algorithm}
\vspace{-10pt}

\section{Numerical Results}\label{sec_sim}
In this section, we provide the numerical simulations to validate the effectiveness of the proposed design. In particular, a MISO interference network with $K$ transmitter-user pairs in a 2D coordinate system is assumed to randomly locate the transmitters and users with $d_{kk}=50~\text{m}$ and $d_{kj}=80~\text{m}$, where $d_{kj}$ represents the distance between transmitter $j$ and user $k$. We adopt the channel model in (\ref{eq_channel_1}), where the numbers of channel paths are assumed to be the same, i.e., $L_{kj}=L,~\forall{k,j}$. The diagonal elements of PRM $\mathbf{\Sigma}_{kj}$ in (\ref{eq_channel_1}) follow the CSCG distribution $\mathcal{CN}\!\left(0,c_{kj}^2/L\right)$, where $c_{kj}^2=\beta_0d_{kj}^{-\alpha_0}$ is the expected channel power gain of $\mathbf{h}_{kj}$, $\beta_0=-40~\text{dB}$ represents the expected value of the average channel power gain at the reference distance of $1$ m, and $\alpha_0=2.8$ denotes the pathloss exponent. It can be pointed out that the total power of the elements in the PRM is the same for the channels with different numbers of paths, i.e., $\mathbb{E}\left\{\operatorname{tr}\left(\mathbf{\Sigma}_{kj}^H\mathbf{\Sigma}_{kj}\right)\right\}\equiv{c_{kj}^2}$, which ensures the fairness of comparison. The elevation and azimuth angles of the channel paths are random variables following the joint probability density function (PDF) $f_A\!\left(\theta_{kj,l},\phi_{kj,l}\right)=\frac{\sin{\theta_{kj,l}}}{2\pi}$, $\theta_{kj,l}\in\left[0,\pi\right]$, $\phi_{kj,l}\in\left[0,\pi\right]$, which indicates that the azimuth angles have the same probability for all directions in the front half-space of antenna array \cite{10318061}. Due to limited scatters in the environment, the $L\times{K}$ elevation and azimuth angles of channel paths from the same transmitter are randomly selected out of the same set of $S=10$ pairs of angles that are randomly generated with the previous joint PDF. Other adopted settings of simulation parameters are $\gamma_{k,\text{min}}=\gamma_{\text{min}}=10~\text{dB}$ and $\sigma_k^2=\sigma^2=-80~\text{dBm}$. The results obtained by \textbf{Algorithm 1} are termed as ``Proposed algorithm". The benchmark schemes are defined as follows. ``MA MRT": the antennas of each transmitter adopting MRT beamforming are deployed at the positions that minimize the total transmit power. ``FPA SOCP": the antennas are fixed at 2D local coordinate systems, respectively, and the SOCP-based beamforming is adopted. ``FPA MRT" can be similarly derived.

\begin{figure}[t]
	\centering
	\includegraphics[width=2.3in]{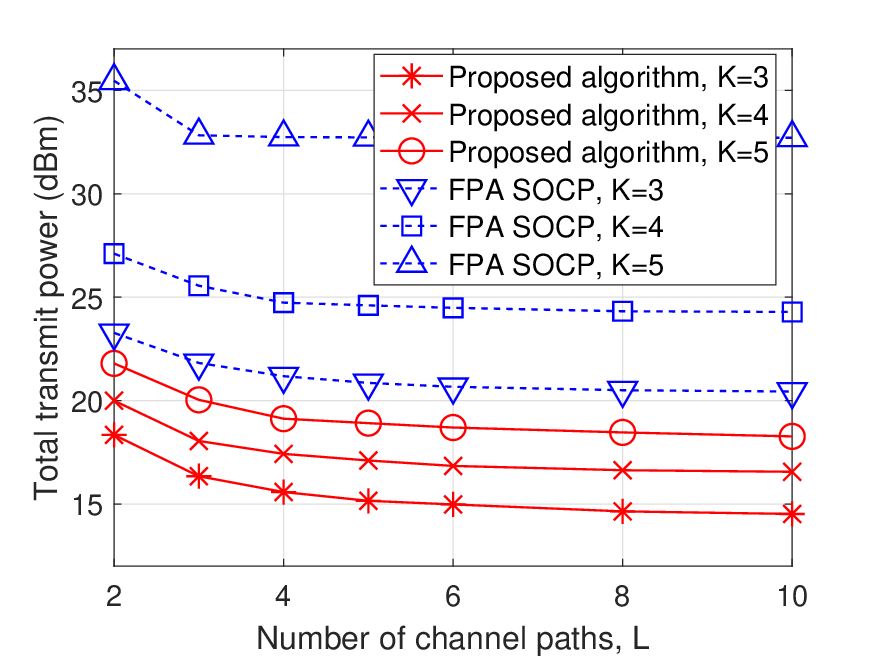}
	\vspace{-3pt}
	\caption{Total transmit power versus the number of channel paths.}
	\label{fig_2}
	\vspace{-10pt}
\end{figure}

First, Fig. \ref{fig_2} shows the total transmit powers versus the numbers of channel paths with different numbers of transmitter-user pairs $K$, where the parameters are set to $A=4\lambda$ and $N=4$. It is observed that the powers of all schemes decrease with $L$ and the proposed algorithm outperforms FPA scheme for any $K$ due to the interference mitigation gain provided by MA positioning optimization. Besides, the decreasing transmit power of FPA system is not caused by the increasing average channel gain (normalized by $L$ and therefore a constant) but due to the reduced interference. As the number of channel paths for each transmitter-user pair increases, the spatial diversity of MA is enhanced by leveraging the prominent channel variation, which decreases the correlation among channel vectors. However, as $L$ is increased larger than $5$, the descent rate for the total transmit power becomes small because of the fact that the channel correlation is constrained by the numbers of elevation and azimuth angles at transmitters. Specifically, according to the channel model in (\ref{eq_channel_1}), if the total numbers of angles are limited, the FRMs of multiple channels are likely to have similar row vectors. Thus, the local movement of MAs cannot further bring a significant reduction in channel correlation. The result demonstrates that due to the strong ability of desired signal enhancement and interference suppression of MA, the total transmit power of the proposed algorithm with $5$ transmitter-user pairs is even lower than that of FPA system with $3$ pairs, which indicates that the MA-aided interference network can accommodate more cells without incurring any increase in total transmit power.

\begin{figure}[t]
	\vspace{-6pt}
	\centering
	\includegraphics[width=2.3in]{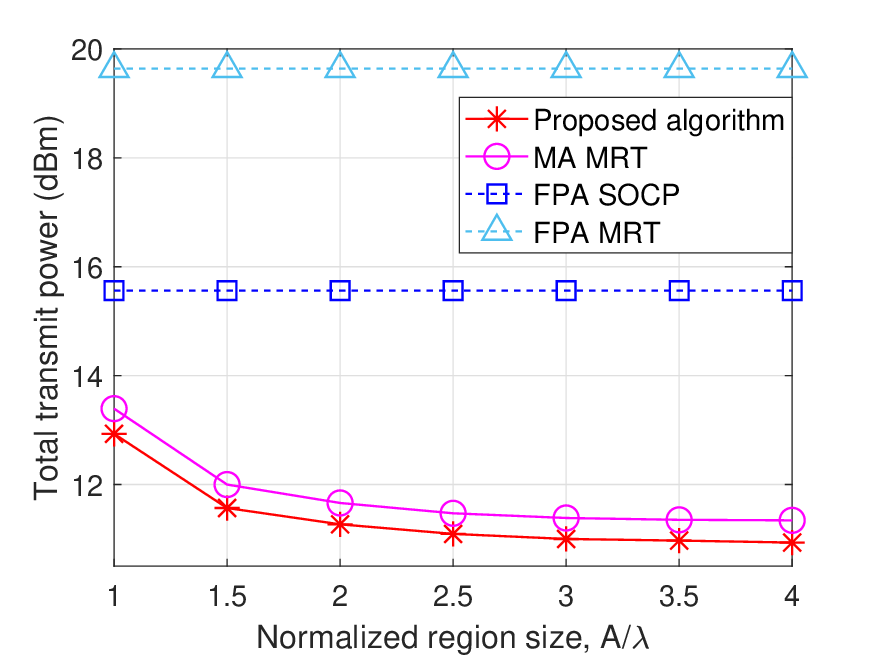}
	\vspace{-5pt}
	\caption{Total transmit power versus the normalized region size.}
	\label{fig_3}
	\vspace{-10pt}
\end{figure}
\begin{figure}[t]
	\centering
	\includegraphics[width=2.3in]{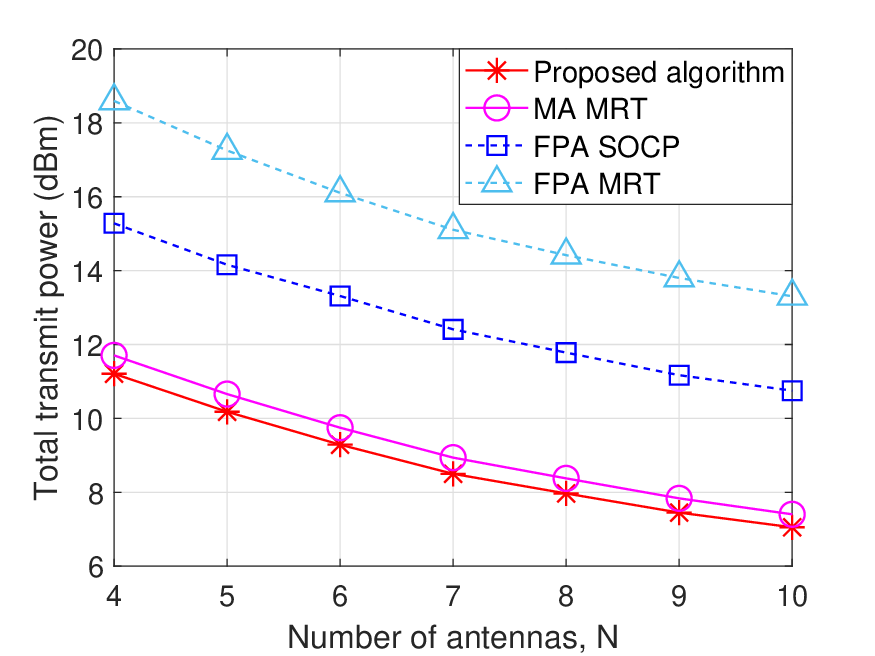}
	\vspace{-5pt}
	\caption{Total transmit power versus the number of antennas.}
	\label{fig_4}
	\vspace{-10pt}
\end{figure}

Fig. \ref{fig_3} shows the total transmit powers of different schemes versus the normalized region sizes for MAs moving at the transmitters, where the size of moving region is normalized by carrier wavelength, i.e., $A/\lambda$. The numbers of transmitter-user pairs, channel paths, and antennas are set as $K=2$, $L=10$, and $N=4$, respectively. From the observation, the total transmit powers achieved by MA schemes decrease with the region size and are much lower than those of FPA schemes. Moreover, the powers of the proposed algorithm and ``MA MRT'' scheme both dramatically decrease as the region size increases from $\lambda$ to $1.5\lambda$, while achieving their lower bounds for $A=3\lambda$. This results from the fact that the larger the region size, the more likely the MAs are to obtain more favorable channel conditions. It is worth noting that as long as the region size is guaranteed to be larger than $\lambda$, the total transmit power of MA system with simple MRT beamforming is significantly lower than that of FPA system, and the power gap becomes larger with the increase of region size. For a relatively large $A$, e.g., $A=2.5\lambda$, the ``MA MRT'' scheme can reap more than $4~\text{dB}$ and $8~\text{dB}$ power-saving compared with the two FPA schemes, respectively.

In Fig. \ref{fig_4}, we compare the total transmit powers of different schemes versus the numbers of antennas, where the parameters are set to $K=2$, $L=10$, and $A=4\lambda$. As can be observed, the total transmit power decreases as the number of antennas increases for all schemes. The superiority of MA schemes over the others, regardless of the type of beamforming employed, becomes evident. This can be attributed to the substantial reduction of the correlation among channel vectors caused by the positioning optimization of MAs, which facilitates effective mitigation of interference among different transmitters and consequently leads to a decrease in total transmit power. It is noted that the utility of SOCP-based beamforming can be approximated by implementing simple MRT method in the MA system, which means that we can drastically reduce the complexity of transmit beamforming with negligible additional power. Besides, the MA scheme is also capable of effectively reducing the number of antennas required by more than half compared with FPA scheme while maintaining the same power constraint and communication metrics. For instance, only $4$ antennas are utilized in ``MA MRT" scheme to achieve the performance of ``FPA SOCP" scheme that deploys $9$ antennas, which simplifies the transmitter design.

\section{Conclusion}\label{sec_conc}
In this paper, we investigated the MA-enabled MISO interference channel system, where each transmitter is equipped with $N$ MAs. By leveraging the additional design DoF provided by MA, we formulated an optimization problem for minimizing the total transmit power of interference network by jointly optimizing the MA positions and transmit beamforming. Since the resultant problem is highly coupled and non-convex, we proposed an alternating optimization algorithm based on the BCD method, where the optimization variables are iteratively updated by introducing the well-designed auxiliary value and invoking the SOCP and SCA techniques. Furthermore, numerical results were provided to clarify that the MA-aided interference network increases the number of cells that can be held and enables the simplification of transmitter design by moving antennas properly within a small region of several-wavelength size.

\bibliographystyle{ieeetr}
\bibliographystyle{IEEEtran}
\bibliographystyle{unsrt}
\bibliography{bibfile}
\end{document}